Critical thickness and orbital ordering in ultrathin La$_{0.7}$Sr$_{0.3}$MnO$_3$ films

M. Huijben,[1,2] L.W. Martin,[1] Y.-H. Chu,[1,3] M.B. Holcomb,[1] P. Yu,[1] G. Rijnders,[2] D.H.A. Blank,[2] and R. Ramesh[1]

[1] *Departments of Materials Science & Engineering and Physics, University of California, Berkeley and Materials Science Division, Lawrence Berkeley National Laboratory, Berkeley, CA 94720, USA.*

[2] *Faculty of Science and Technology and MESA+ Institute for Nanotechnology, University of Twente, P.O. BOX 217, 7500 AE, Enschede, The Netherlands.*

[3] *Department of Materials Science and Engineering, National Chiao Tung University, Hsinchu 30010, Taiwan.*

Detailed analysis of transport, magnetism and x-ray absorption spectroscopy measurements on ultrathin La$_{0.7}$Sr$_{0.3}$MnO$_3$ films with thicknesses from 3 to 70 unit cells resulted in the identification of a lower critical thickness for a non-metallic, non-ferromagnetic layer at the interface with the SrTiO$_3$ (001) substrate of only 3 unit cells (~12 Å). Furthermore, linear dichroism measurements demonstrate the presence of a preferred ($x^2$-$y^2$) in-plane orbital ordering for all layer thicknesses without any orbital reconstruction at the interface. A crucial requirement for the accurate study of these ultrathin films is a controlled growth process, offering the coexistence of layer-by-layer growth and bulk-like magnetic/transport properties.

I. INTRODUCTION

Doped-manganite perovskites have been extensively investigated in the last decade due to their tantalizing magnetotransport properties, such as colossal magnetoresistance[1] and half-metallicity[2], which make them very promising for applications in spintronic devices[3–8]. The properties of manganites are related to the coupling of the lattice, orbital, charge and spin, which can be affected in various ways: ionic radii[9], lattice strain[10–12], doping level[13], oxygen stoichiometry[14], temperature, magnetic field[15], electric field[16], hydrostatic pressure[17,18] and photoexcitation[19]. Many doped-manganites exhibit a magnetic transition at the Curie temperature, $T_C$, accompanied by a metal-insulator transition, which makes them paramagnetic insulators at high temperature and ferromagnetic half-metals at low temperature[9,20–23].

Many investigations have focused on $La_{0.7}Sr_{0.3}MnO_3$ (LSMO), which shows the highest Curie temperature among the family of manganites ($T_C$ ~369 K)[13]. An extensive number of studies have demonstrated the fabrication of such manganites in the form of thin films, which display very different properties as compared to bulk. Lattice mismatch can cause structural modifications at the interface between the film and the substrate, strongly affecting the magnetic properties[24]. For example, tensile strain suppresses ferromagnetism and reduces the ferromagnetic Curie temperature, which is generally interpreted by considering a strain-induced distortion of $MnO_6$ octahedra based on Jahn-Teller distortion theory[25]. The existence of an interfacial layer, which shows insulating behavior over the whole temperature range without a metal-insulator transition, has been suggested. A so-called 'dead-layer' thickness can be defined as the thinnest layer for which metallic as well as ferromagnetic behavior is observed. Estimates of a dead layer thickness for films on $NdGaO_3$ (110), $SrTiO_3$ (001), MgO (001) and $LaAlO_3$ (001) substrates are ~ 3, 4, 4 and 5 nm, respectively[26–28]. However, the origin of the dead layer is still controversial. Magnetic-resonance measurements[29,30] and scanning tunneling spectroscopy[31] have shown that it is connected to a phase-separation phenomenon[32] at the interface where ferromagnetic insulating and metallic phases separate at a scale of a few nanometers[33]. It has been suggested that the mechanism driving such a phase separation is related to the presence of structural inhomogeneities localized at the interface between film and substrate. Very recently, evidence was also given for a possible orbital reconstruction at the LSMO interface, for which a strain-induced distortion of the $MnO_6$ octahedra led to crystal field splitting of the $e_g$ levels and lowering the $(3z^2-r^2)$ orbital over the $(x^2-y^2)$ orbital[34] resulting in a local C-type antiferromagnetic structure at the interface.

Although the material properties of LSMO films depend strongly on growth mode and oxidation level, most of the studies on ultrathin LSMO films commonly focused on the optimization of only one of them. Typically, such focusing on growth behavior can lead to inferior magnetic properties. In this report, special attention was given to the relationship between growth mode, oxidation level and material properties to ensure the coexistence of layer-by-layer growth and bulk-like magnetic/transport properties. Both are required to study accurately the behavior of ultrathin LSMO films. In order to obtain information about the suggested presence of dead layers and orbital reconstruction at LSMO interfaces, we studied high-quality ultrathin LSMO films down to a thickness of 3 unit cells (~12 Å). $SrTiO_3$ (001) single crystals were selected as substrates due to the low mismatch with LSMO and the fact that the surface can be well defined by a single termination treatment. The growth dynamics were investigated by monitoring the intensity variations of various features in the reflection high energy electron diffraction (RHEED) patterns. Transport and magnetic properties of the ultrathin films were measured to provide individual critical thickness values for metallic and ferromagnetic behavior, which together were used to determine the so-called 'dead layer' thickness. Furthermore, linear dichroism (LD) was investigated by x-ray absorption spectroscopy (XAS) to demonstrate the absence of orbital reconstruction for ultrathin LSMO films.

II. EXPERIMENT

The films used in this study were grown epitaxially using pulsed laser deposition from a stoichiometric $La_{0.7}Sr_{0.3}MnO_3$ target on $TiO_2$-terminated $SrTiO_3$ (100) substrates[35] by applying a KrF excimer laser at a repetition rate of 1 Hz and a laser fluence of ~1.5 J cm$^{-2}$. During growth, the substrate was held at 750 °C in an oxygen environment in the range 100-300 mTorr. The growth was monitored in-situ by RHEED analysis[36], allowing precise control of the thickness at the unit cell (u.c.) scale and accurate characterization of the growth dynamics. For these conditions a growth rate was determined of ~0.12 Å/sec. After the growth, the samples were slowly cooled to room temperature in 1 atm. of oxygen at a rate of ~5 °C/min to improve the oxidation level.

RHEED has already proven to be a very versatile technique for growth and surface studies of thin films and has also been used to investigate the growth of $La_{1-x}Sr_xMnO_3$ thin films[24,37]. Although layer-by-layer growth was concluded in earlier studies from the observed intensity oscillations, interruption of the deposition was then still necessary to recover them. For our growth conditions such interruptions aren't necessary anymore and continuous layer-by-layer growth can be maintained. On the other hand more attention has been given to the dependence of the growth mode on the deposition pressure and additional information from the features in the corresponding RHEED diffraction patterns has been utilized in contrast to previous studies. Analysis techniques, such as atomic force microscopy (AFM) and four-circle x-ray diffraction (XRD), were additionally used to demonstrate the atomically smooth surface and single crystal structure (not shown in this report).

Transport properties were determined in a Van der Pauw four-probe configuration with a Quantum Design physical properties (PPMS) measurement system in the range 10-350 K. Gold contacts had to be deposited to establish good ohmic contact between aluminum bonding wires and the LSMO surface. A Quantum Design superconducting quantum interference device (SQUID) measurement system was used to measure the magnetic properties in the temperature range 10-380 K with the magnetic field applied in-plane along the (100)-direction of the SrTiO$_3$ crystal. Linear dichroism was investigated for some LSMO films by x-ray absorption spectroscopy measurements of the Mn 2$p$ edge at beamline 4.0.2 of the Advanced Light Source[38] (Lawrence Berkeley National Laboratory) to directly probe the orbital character of 3$d$ states in these ultrathin manganite films[39].

III. RESULTS

A. Unit cell controlled growth process

Figure 1 shows the RHEED specular intensities recorded during the initial growth of the first La$_{0.7}$Sr$_{0.3}$MnO$_3$ unit cells on TiO$_2$-terminated SrTiO$_3$ surfaces at various oxygen deposition pressures of 100, 200 and 300 mTorr, respectively. Clear RHEED oscillations can be observed, indicating layer-by-layer growth. However, large differences in the RHEED intensities for the different pressures indicate the possible occurrence of different growth modes. For LSMO growth at 100 and 200 mTorr (Fig.1(a) and 1(b)) RHEED oscillations will remain visible for the total deposition of 70 unit cells (~28 nm), suggesting continuous layer-by-layer growth. No additional interruption of the deposition was necessary to recover them as demonstrated by earlier studies[24]. The absence of islands is confirmed by the presence of sharp 2-dimensional spots in the corresponding RHEED diffraction patterns of the surface of the 70 unit cells thick layer, which are shown in the insets. The presence of these sharp spots lying on concentric Laue circles indicates true reflective diffraction from a smooth surface. However, for LSMO growth at 300 mTorr RHEED oscillations disappear very quickly after ~10 unit cells and the growth mode changes from layer-by-layer growth to island growth. This is confirmed by the periodic pattern of vague 3-dimensional spots in the RHEED pattern, which originates from transmission through particles on the surface. These results were confirmed by atomic force microscopy, displaying island formation for growth at 300 mTorr and atomically smooth terraces separated by unit cell steps, similar to the original substrate surface, for deposition pressures of 100 and 200 mTorr.

The magnetic properties of the LSMO films were measured in a superconducting quantum interference device (SQUID) system. Figure 2(a) shows clear ferromagnetic hysteresis loops for the different samples at 10 K after magnetic field cooling at 1 Tesla from 360 K. However, large differences in the saturation magnetization values can be observed. LSMO films grown at 200 and 300 mTorr display a magnetization of ~550 emu/cm$^3$ comparable to bulk values, while films grown at 100 mTorr exhibit a reduced magnetization of ~300 emu/cm$^3$. These results together with the RHEED observations indicate that an oxygen deficiency is responsible for the reduced magnetization and not the surface roughness. Although LSMO growth at 100 mTorr shows the best RHEED oscillations and a smooth surface, it possesses inferior magnetic properties when compared to films grown at 200 mTorr. The diminished quality of the magnetic properties is also demonstrated by the increase of the coercive field from ~12 Oe to ~40 Oe, see inset Fig. 2(a), and the decrease of the Curie temperature, $T_C$, from ~345 K to ~330 K, see Fig. 2(b). Annealing the LSMO films after growth in 1 atm. oxygen for 60 min. at 750 $^o$C and during subsequent cool down did not improve the magnetic properties, as can be seen in Figure 2. From these results we can conclude that high-quality unit cell controlled LSMO layers with bulk-like magnetic properties can be fabricated in a narrow deposition regime, but only when the growth dynamics and magnetic properties are monitored simultaneously and not focusing on only one individual part.

Subsequently, thin films of LSMO with various thicknesses between 3 and 70 unit cells were grown at the optimum deposition temperature and oxygen pressure, 750 °C and 200 mTorr, to ensure the coexistence of layer-by-layer growth and bulk-like magnetic properties. Figure 3 shows clear RHEED intensity oscillations during the total growth of the various samples indicating controlled layer-by-layer growth of individual unit cells. X-ray diffraction analysis of the thickest samples revealed that they are tensile strained on the SrTiO$_3$ (001) substrate (~3.905 Å) and have a shortening of the c-axis (~3.843 Å) as compared to bulk (~3.889 Å)[40].

B. Transport measurements

The temperature dependence of the resisitivity in zero magnetic field for ultrathin LSMO films with variable thickness between 3 and 70 unit cells is given in Fig. 4(a). Thicker films show a bulk-like metallic behavior over the whole temperature regime. The residual resistivity at 10 K is in the 60-80 μΩ cm range, which is close to previously reported values[26]. When the LSMO layer thickness decreases the resistivity increases quite drastically over the whole temperature range. For LSMO layers with thicknesses of 8, 5 and 3 unit cells a continuous resistivity increase is observed with a pronounced rise, which indicates the presence of a layer with dramatically reduced conductivity. This can be seen more clearly in the thickness dependence of the total conductance G of the films, where G = 1/R$_S$ with R$_S$ as the measured sheet resistance (Fig. 4(b)). A linear thickness dependence of the conductance was found as expected for uniform films. The intercept of the horizontal axis provides an estimate of the non-metallic layer of about 8 unit cells (~32 Å), which is lower than the previously reported critical thickness of metallic behavior (~40 Å)[28] for LSMO on SrTiO$_3$ (001).

C. Magnetic measurements

The magnetic properties of the ultrathin LSMO films are shown in Figure 5. A large reduction of the saturation magnetization as well as an increase of the coercive field H$_C$ (Fig. 5(a)) can be observed for thicknesses below 13 unit cells (~48 Å). At the same time the Curie temperature T$_C$ is lowered as well, displaying a double transition for a layer thickness of 8 unit cells (~32 Å) before drastically decreasing for LSMO films of only 5 and 3 unit cells (~20 Å & ~12 Å, respectively) (Fig. 5(b)). The thickness dependence has been summarized in Fig. 5(c), where coercive field and Curie temperature, H$_C$ and T$_C$, are nearly constant for thicknesses down to 13 unit cells. Further reduction of the layer thickness results in a dramatic change in the magnetic properties, although the films remain ferromagnetic down to 3 unit cells (~12 Å). The observation of a critical thickness for ferromagnetism of 3 unit cells (~12 Å) together with the measured critical thickness for metallicity of 8 unit cells (~32 Å) results in the determination of a 'dead layer' thickness of 8 unit cells, which is the thinnest layer with metallic as well as ferromagnetic behavior. This new lower value for a so-called 'dead-layer' of ~32 Å is about 2 unit cells lower then the previously reported 'deadlayer' thickness (~40 Å)[28] for LSMO on SrTiO$_3$ (001).

D. XAS measurements

To study the orbital ordering of the 3*d* states in these ultrathin LSMO films, XAS spectra were recorded at the Mn-*L*$_{2,3}$ edge corresponding to the 2*p* → 3*d* resonant transition. The spin-orbit interaction of the Mn 2$p$ core hole splits the spectrum into two broad multiplets, the $L_3$ (2$p_{3/2}$) edge at lower photon energy and the $L_2$ (2$p_{1/2}$) edge at higher photon energy. Using linearly polarized radiation, two spectra can be measured when the polarization vector is set parallel to the $c$ crystallographic axis or perpendicular to it ($I_c$ and $I_{ab}$, respectively). The difference between those spectra ($I_{ab}$ - $I_c$) provides the linear dichroism (LD) values and gives direct insight of the empty Mn 3$d$ states[39]. Considering the crystal field splitting, the effect can be mainly related to the occupation of the two $e_g$ states (3$z^2$-$r^2$) and ($x^2$-$y^2$) with majority spin: a LD which is on average positive (negative) is due to a preferential occupation of the in-plane ($x^2$-$y^2$) (out-of-plane (3$z^2$-$r^2$)) orbital.

LD-XAS measurements were performed on ultrathin LSMO films on SrTiO$_3$ (001) substrates with LSMO layer thicknesses of 5, 8 and 70 unit cells, as described above. The experimental spectra of the ultrathin films are shown in Fig. 6(a) and, although the intensity changes with layer thickness, the shape remains essentially unchanged. When the LD signal is investigated in detail, see Fig. 6(b), small changes in relative LD signal intensity and shape can be observed. However, the average LD signal remains positive for all three LSMO thicknesses, which means that the preferred orbital ordering stays ($x^2$-$y^2$) in-plane for all LSMO films down to layer thicknesses of 5 unit cells. This result is in sharp contrast with earlier observations of orbital reconstruction in ultrathin LSMO films[34] and can even be strengthened by the absence of a sign reversal for a photon energy directly above E ~ 644 eV, which was used as an additional qualitative argument by Tebano et al.

## III. DISCUSSION

The results of transport, magnetic and x-ray absorption spectroscopy measurements all indicate a gradual change in magnetoelectronic structure for our ultrathin LSMO films. In this section we will discuss the observations of a reduced thickness for a non-metallic, non-ferromagnetic interface layer, as well as the absence of any orbital reconstruction, with respect to the unit-cell controlled growth process.

To study the properties of ultrathin films correctly, the first important task was to determine a fabrication procedure which could deliver bulk-like magnetic/transport properties, while still providing a growth control on the unit cell scale. Since the magnetic/transport properties depend strongly on the oxygen stoichiometry, special attention had to be given to the relation between growth dynamics and oxidation level. Although we analyzed the RHEED intensity variations as other groups have done before, more attention was given to specific shapes in the corresponding RHEED patterns. Together with the measurements of the magnetic (transport) properties in a SQUID (PPMS) system, we determined that ideal, atomically smooth LSMO layers with bulk-like properties could only be fabricated in a small regime with deposition conditions of 750 $^{o}$C and 200 mTorr. Focus on either the growth dynamics or the magnetic/transport properties would result in inferior properties of the other. However, such specific focussing was commonly done in previous studies of ultrathin LSMO films and we believe that the introduction of detailed analysis of the growth dynamics in our study provides the best method to accurately investigate the properties of only a few unit cells of LSMO.

Detailed investigations of ultrathin LSMO films, where the thicknesses were controlled at the unit cell scale, demonstrated clearly the existence of an non-metallic, nonferromagnetic layer at the interface between the LSMO layer and the SrTiO$_3$ substrate. Individually, critical thickness values were determined for metallicity of 8 unit cells (~32 Å) and ferromagnetism of 3 unit cells (~12 Å). Interestingly, very recently an elongation of the out-of-plane lattice constant at the substrate-film interface was determined by surface x-ray diffraction[41] with a critical thickness of precisely 3 unit cells. This increase in c-axis at the interface is the opposite of the expected decrease for in-plane tensile strained layers. No interfacial defects could be revealed and, therefore, the role of a possible interface roughness was negligible. They concluded that stoichiometry changes were responsible for the dilation at the substrate-film interface with most likely Sr enrichment in the topmost monolayer, due to segregation. However, such local elongations of the crystal structure have also been determined by surface x-ray diffraction at the interface[42,43] between $LaAlO_3$ and $SrTiO_3$. In that case the $LaAlO_3$ was also tensile strained to the $SrTiO_3$ substrate and similarly resulted in an unexpected dilation of the first few unit cells at the interface. Displacements of the cations and anions, due to electronic reconstruction at this polar/non-polar interface, were suggested to cause a 2-dimensional conducting layer at the interface between those otherwise insulating oxides. For our interface between the LSMO layer and the $SrTiO_3$ substrate a dramatic change in material properties is very conceivable without any variations in the stoichiometry, but only caused by very small changes in the atomic positions close to the interface. However, more detailed experiments are necessary to determine if structural or stoichiometric changes play an important role.

Furthermore, linear dichroism analysis from XAS measurements have been used to study the orbital ordering for several LSMO films with various thicknesses. The preferred orbital ordering was ($x^2$-$y^2$) in-plane for all layer thicknesses down to only 5 unit cells. This is in good agreement with previous studies of relatively thick LSMO films (100 u.c.) on $SrTiO_3$ (001) substrates[37], which gave evidence of a preferential occupation of the in-plane ($x^2$-$y^2$) orbitals as a consequence of the decrease of the in-plane Mn-O bond lengths due to tensile epitaxial strain. However, in our case no change in orbital ordering was found also for ultrathin LSMO films and, therefore, the suggestion of orbital reconstruction at the interface[34] was denounced. An out-of-plane ($3z^2$-$r^2$) orbital ordering, as observed also in LSMO layers on

LaAlO$_3$ (001) substrates[37,44], indicating a C-type antiferromagnetic insulating phase was not present in our ultrathin LSMO films down to a thickness of only 5 unit cells.

## IV. CONCLUSION

We have studied ultrathin LSMO films with thicknesses ranging from 3 to 70 unit cells on SrTiO$_3$ substrates. It is found that an ideal unit cell controlled growth process, resulting in layer-by-layer growth and bulk-like magnetic/transport properties, can only be achieved in a small regime with growth conditions of 750 °C and 200 mTorr. By using this optimized growth process, we have been able to study the properties of LSMO layers with thicknesses down to only a few unit cells. As a result, new lower values have been determined for the critical LSMO layer thickness to produce ferromagnetism (~12 Å) and metallicity (~32 Å). Although the magnetic ordering and conductivity behavior are coupled in LSMO layers, a large difference was found for their critical thicknesses. An explanation could be the loss of a conducting percolation path for LSMO layers thinner than ~32 Å, which could possibly be caused by phase separation due to the formation of ferromagnetic/metallic and non-ferromagnetic/non-metallic regions. This effect has been predicted in a recent theoretical study[45], which modeled the changes in average ionic charge per atomic (001) plane. They found that an electronic phase separation between a ferromagnetic/metallic phase and a spin- and orbital-ordered insulator phase occurs at the manganite-insulator interface. This instability is favored by the reduction of carriers at the interface, which weakens the ferromagnetic coupling between the Mn ions, making more relevant the superexchange antiferromagnetic interaction. Interestingly, our new value of ~12 Å for the critical LSMO layer thickness, which is completely non-ferromagnetic as well as nonmetallic, matches nicely with previous observations by surface x-ray diffraction of an elongated crystal structure at the film-substrate interface as well as previous theoretical modeling of phase separation locally at the manganite-insulator interface.

Detailed analysis of linear dichroism by x-ray absorption spectroscopy in these ultrathin LSMO films has also demonstrated that the preferred orbital ordering remains ($x^2$-$y^2$) in-plane for all layer thicknesses, which clearly opposes previous reports of an orbital reconstruction at the interface into out-of-plane ($3z^2$-$r^2$) orbitals. In total we can conclude that heterostructures with high quality LSMO films can be fabricated for device applications, even with thicknesses down to several unit cells. However, dramatic changes in the material properties will always be significant locally at the interfaces.

Acknowledgements

This work was supported by the Director, Office of Science, Office of Basic Energy Sciences, Materials Sciences and Engineering Division of the US Department of Energy under Contract No. DE-AC02-05CH11231. G.R. acknowledges support by the Netherlands Organization for Scientific Research (NWO).

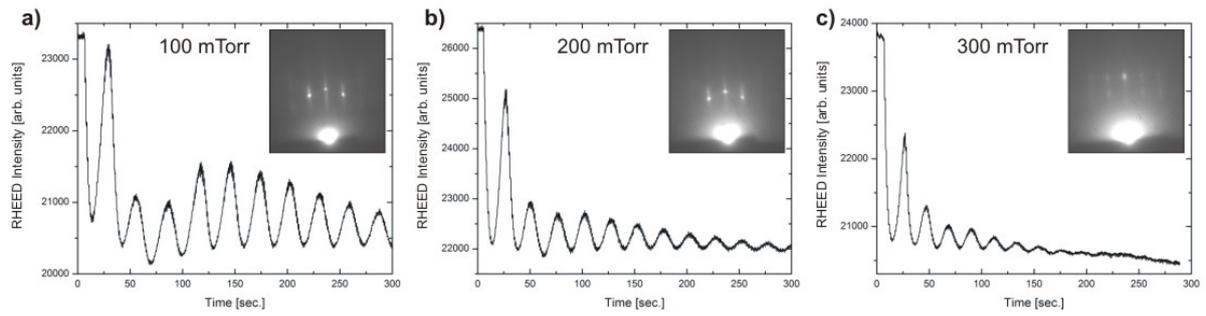

**Figure 1**: Surface analysis by reflection high-energy electron diffraction (RHEED) during initial growth of $La_{0.7}Sr_{0.3}MnO_3$ on $TiO_2$-terminated $SrTiO_3$ (001) substrates at various oxygen deposition pressures of 100 mTorr (a), 200 mTorr (b) and 300 mTorr (c). The insets display the RHEED diffraction patterns after the growth of a 70 unit cells (~28 nm) thick $La_{0.7}Sr_{0.3}MnO_3$ layer.

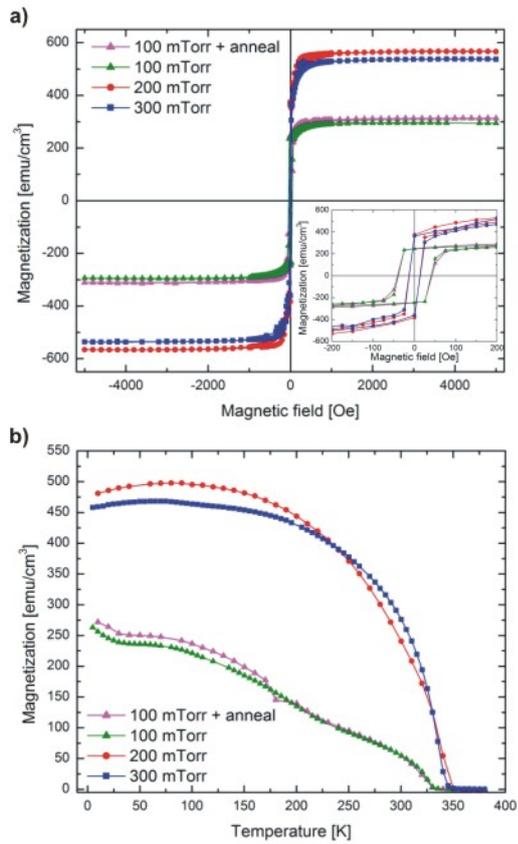

**Figure 2**: Magnetic properties of 28 nm thick $La_{0.7}Sr_{0.3}MnO_3$ films on $SrTiO_3$ (001) grown at different oxygen pressures. (a) Magnetic hysteresis loops measured at 10 K. The diamagnetic contribution to magnetization (not shown) has been attributed to the substrate and has been subtracted. The inset shows an enlargement near the origin. (b) Temperature dependence of the magnetization measured at 100 Oe. All samples were field cooled at 1 Tesla from 360 K along the [100] direction before the measurements were performed.

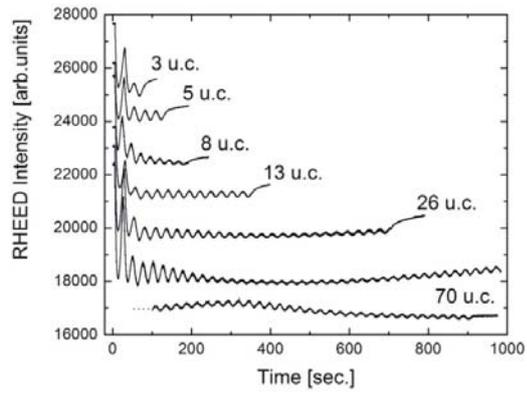

**Figure 3**: RHEED intensity recorded during growth of $La_{0.7}Sr_{0.3}MnO_3$ ultrathin films on $TiO_2$-terminated $SrTiO_3$ (001) with thicknesses from 3 to 70 unit cells at the optimum deposition temperature and oxygen pressure of 750 °C and 200 mTorr, respectively.

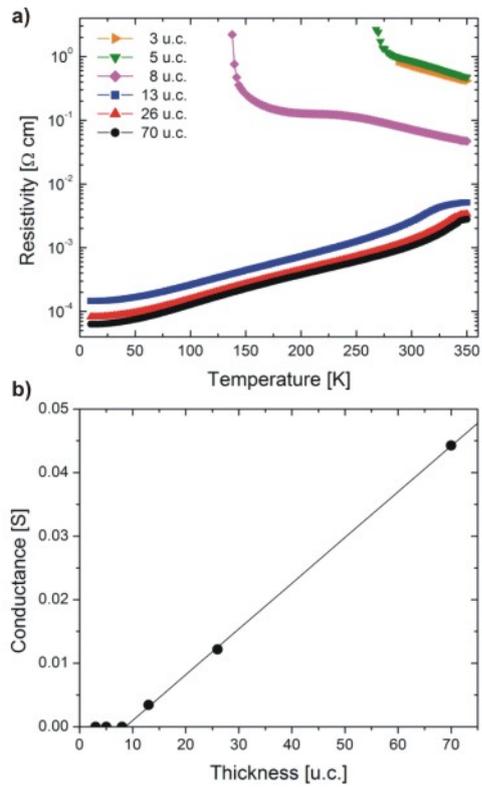

**Figure 4**: Transport properties of ultrathin LSMO films on SrTiO$_3$ (001) substrates. (a) Temperature-dependent resistivity for films of different thicknesses. (b) Thickness dependence of the total conductance of films at 10 K.

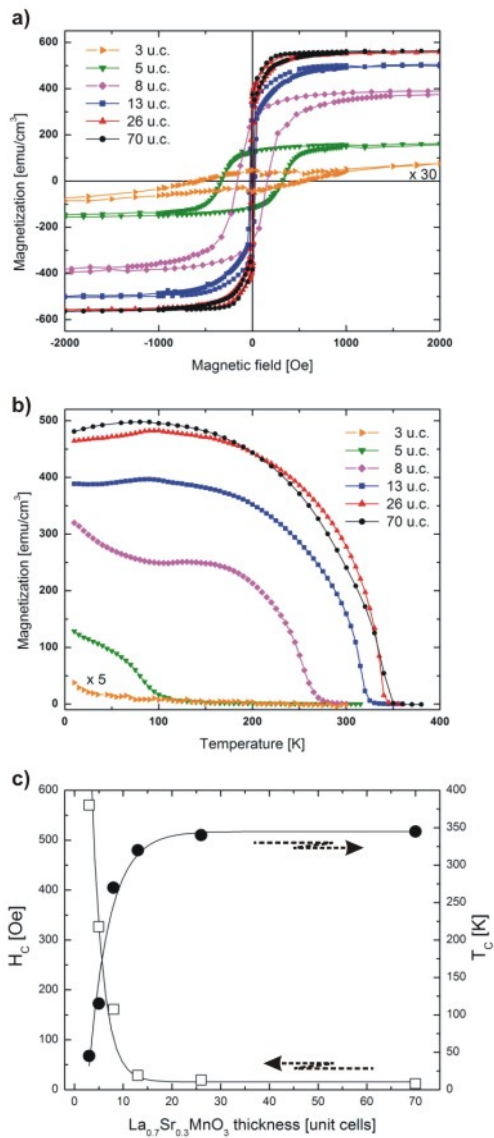

**Figure 5**: Ferromagnetic properties of ultrathin LSMO films on SrTiO$_3$ (001). (a) Magnetic hysteresis loops measured at 10 K. The diamagnetic contribution to magnetization (not shown) has been attributed to the substrate and has been subtracted. (b) Temperature dependence of the magnetization measured at 100 Oe. All samples were field cooled at 1 Tesla from 360 K along the [100] direction before the measurements were performed. (c) Layer thickness dependence on the coercive field $H_C$ and the Curie temperature $T_C$.

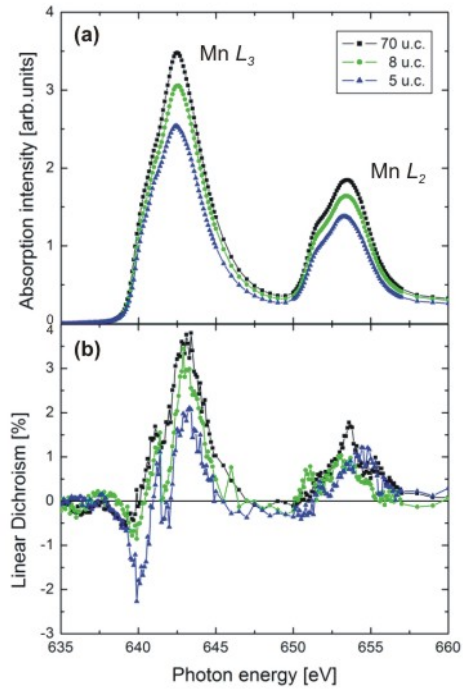

**Figure 6**: Experimental XAS and LD measurements at 100 K at the Mn-$L_{2,3}$ edge for LSMO films with thicknesses of 5, 8 and 70 unit cells: (a) average between XAS signals taken in both polarization directions; (b) LD in percent of the XAS $L_3$ peak height signal.